\documentclass[12pt]{article}

\usepackage[utf8x]{inputenc}      % input font encoding
\usepackage{hyperref}
\usepackage{amsmath,amssymb,mathtools}
\usepackage{cleveref}
\usepackage[T1]{fontenc}          % output font encoding
\usepackage{booktabs,tabularx}
\usepackage{graphicx,subfig}
\usepackage{xspace}
\usepackage[usenames]{xcolor}\definecolor{fscolor}{RGB}{44,118,255}
\usepackage{geometry}
\usepackage{todonotes}
\usepackage{listings}
\usepackage[absolute]{textpos}
\usepackage[many]{tcolorbox}
\usepackage{xparse}
\usepackage[font=small,labelfont=bf,format=plain,margin=0.05\textwidth]{caption}
\usepackage{bbm}
\usepackage{tabularx}
\usepackage{cite}
\usepackage{soul}

\crefname{figure}{Fig.}{Figs.}
\crefname{equation}{Eq.}{Eqs.}
\crefname{section}{Sec.}{Secs.}

\allowdisplaybreaks

\evensidemargin \oddsidemargin
\newcommand{\DR}{{\ensuremath{\overline{\text{DR}}}}\xspace}
\newcommand{\MDR}{{\ensuremath{\overline{\text{MDR}}}}\xspace}
\newcommand{\MS}{{\ensuremath{\overline{\text{MS}}}}\xspace}

\newcommand{\MSSM}{{\text{MSSM}}}

\newcommand{\SUSY}{{\text{SUSY}}}

\newcommand{\FH}{\mbox{{\tt FeynHiggs}}\xspace}

\newcommand{\Sec}[1]{Section~\ref{#1}}
\newcommand{\App}[1]{App.~\ref{#1}}

\newcommand{\cp}{\ensuremath{{\cal CP}}}

\newcommand{\msusy}{\ensuremath{M_\SUSY}\xspace}

\newcommand{\mgls}{\ensuremath{\vert M_3\vert}\xspace}

\newcommand{\xt}{\ensuremath{\widehat X_t}\xspace}

\newcommand{\gev}{\,\, \mathrm{GeV}}

\newcommand{\order}[1]{\ensuremath{{\cal O}(#1)}}
\newcommand{\al}{\alpha}
\newcommand{\als}{\al_s}
\newcommand{\alt}{\al_t}

\begin{document}

\thispagestyle{empty}
\def\thefootnote{\fnsymbol{footnote}}

\begin{flushright}
DESY 19-235
\end{flushright}
\vspace{3em}
\begin{center}
{\Large\bf
Precise prediction for the mass of the light MSSM\\[.3em]
Higgs boson for the case of a heavy gluino
%Avoiding large gluino contributions in the calculation of the light MSSM
%Higgs boson mass
}
\\
\vspace{3em}
{
Henning Bahl\footnote{email: henning.bahl@desy.de},
Ivan Sobolev\footnote{email: ivan.sobolev@desy.de},
Georg Weiglein\footnote{email: georg.weiglein@desy.de}
}\\[2em]
{\sl Deutsches Elektronen-Synchrotron DESY, Notkestra{\ss}e 85, D-22607 Hamburg, Germany}
\def\thefootnote{\arabic{footnote}}
\setcounter{page}{0}
\setcounter{footnote}{0}
\end{center}
\vspace{2ex}
\begin{abstract}
{}

State-of-the-art predictions for the mass of the lightest MSSM Higgs boson usually involve the resummation of higher-order logarithmic contributions obtained within an effective-field-theory (EFT) approach, often combined with a fixed-order calculation into a hybrid result. For the phenomenologically interesting case of a significant hierarchy between the gluino mass and the masses of the scalar top quarks the predictions suffer from large theoretical uncertainties related to non-decoupling power-enhanced gluino contributions in the EFT results employing the \DR renormalisation scheme. We demonstrate that the theoretical predictions in the heavy gluino region are vastly improved by the introduction of a suitable renormalisation scheme for the EFT calculation. It is shown that within this scheme a recently proposed resummation of large gluino contributions is absorbed into the model parameters, resulting in reliable and numerically stable predictions in the heavy-gluino region. We also discuss the integration of the results into the public code \FH.

\end{abstract}

\newpage
\def\thefootnote{\arabic{footnote}}

%%%%%%%%%%%%%%%%%%%%%%%%%%%%%%%%%%%%%%%%%%%%%%%%%%%%%%%%%%%%%%%%%%%%%%%%%%%%%%
%%%%%%%%%%%%%%%%%%%%%%%%%%%%%%%%%%%%%%%%%%%%%%%%%%%%%%%%%%%%%%%%%%%%%%%%%%%%%%

\section{Introduction}
\label{sec:01_intro}

In models that allow a direct prediction of the mass of the Higgs boson resembling the one of the Standard Model (SM) in terms of the model parameters important constraints on the parameter space can be obtained from the comparison of the predicted value with the precise mass measurement of the detected Higgs signal~\cite{Aad:2012tfa,Chatrchyan:2012xdj,Aad:2015zhl}. The most thoroughly studied model in this context is the Minimal Supersymmetric extension of the Standard Model (MSSM). The accuracy of the prediction for the light \cp-even Higgs-boson mass of the MSSM has been significantly improved during the last years (see~\cite{Borowka:2018anu,Bahl:2018jom,Harlander:2018yhj,Bahl:2018ykj, R.:2019ply,Bagnaschi:2019esc,R.:2019irs,Murphy:2019qpm,Goodsell:2019zfs, Harlander:2019dge,Bahl:2019hmm} for recent works), in particular for the case where some or all of the supersymmetric (SUSY) particles are relatively heavy. This has been achieved with effective-field-theory (EFT) methods, where the heavy SUSY particles are integrated out. The contributions obtained in this way employ the \MS and \DR renormalisation schemes for the running from the low scale to the SUSY scale(s) and the matching with the full theory at the high scale. Hybrid results~\cite{Hahn:2013ria,Bahl:2016brp,Athron:2016fuq,Staub:2017jnp,Bahl:2017aev,Athron:2017fvs,Bahl:2018jom,Bahl:2018ykj,Harlander:2019dge}, combining a fixed-order calculation with an EFT treatment for resumming higher-order logarithmic contributions, provide accurate predictions both for low and high values of the SUSY scale.

Appropriate EFT descriptions have been developed for different patterns of possible SUSY spectra, including split-SUSY type scenarios where the mass of the gluino --- the superpartner of the gluon --- is much lighter than the masses of the scalar quarks~\cite{Giudice:2011cg,Bagnaschi:2014rsa}. However, no proper treatment of the case where the gluino is significantly heavier than the scalar quarks is available up to now. As a consequence, large theoretical uncertainties in the Higgs-mass prediction occur for values of $\mgls \gtrsim 2 \msusy$~\cite{Bahl:2019hmm}, where $M_3$ denotes the gluino mass parameter, and $\msusy$ denotes the geometric mean of the soft-SUSY breaking masses of the stops, the superpartners of the top quark. This is a serious drawback for realistic analyses of SUSY phenomenology since the LHC searches have pushed the experimental bounds on the gluino mass to the region above $\sim 2$ TeV, while the superpartners of the top quark are still allowed to have masses around the TeV scale~\cite{Aaboud:2017vwy,Aad:2019ftg,ATLAS-CONF-2019-040,Sirunyan:2017pjw, Sirunyan:2018vjp,Sirunyan:2019ctn,Sirunyan:2019xwh,Sirunyan:2019zfh,CMS-PAS-SUS-19-008}.

The large theoretical uncertainties for the case where the gluino is heavier than the stop particles can be traced to corrections to the squared masses of the stops that are proportional to $\mgls^2$ at the one-loop level in the \DR scheme as well as corrections linear in \mgls originating from one-loop corrections to the stop mixing parameter $X_t$. If instead an on-shell (OS) renormalisation for the stop masses and the stop mixing parameter (it is sufficient in this context to impose a condition on the renormalised off-diagonal self-energy of the two scalar top quarks) is employed, the momentum subtraction arising from the on-shell counterterms leads to a cancellation of the leading contributions that are proportional to $\mgls^2$ and $\mgls$. Accordingly, the two-loop fixed-order prediction for the mass of the SM-like Higgs boson of the MSSM in the OS scheme depends only logarithmically on the gluino mass~\cite{Heinemeyer:1998np,Degrassi:2001yf,Borowka:2014wla}, while the corresponding \DR result contains contributions that are enhanced by powers of $\mgls$~\cite{Degrassi:2001yf}. However, for EFT calculations the OS scheme is not applicable. As a consequence, large non-decoupling effects for a heavy gluino occur both in pure EFT results (using a single SUSY scale) and also in the EFT parts of hybrid results via the threshold corrections at the SUSY scale that are evaluated in the \DR scheme.

A possible solution would be the derivation of a complete EFT where the effects of a heavy gluino are systematically integrated out from the MSSM. While in the context of other observables such an approach has been investigated~\cite{Muhlleitner:2008yw,Aebischer:2017aqa,Kramer:2019fwz}, a complete EFT calculation for a heavy gluino that could be applied for the Higgs-mass prediction in the MSSM has not been carried out so far. In~\cite{Vega:2015fna} it was proposed to deal with this problem by reexpressing the threshold corrections in a pure (single-scale) EFT result derived in the \DR scheme in terms of an ``OS-like'' renormalisation scheme. However, this prescription is not a viable option since its derivation was based upon an incorrect result for the transition of the OS stop mixing parameter to the \DR stop mixing parameter. If the correct formula is used, a large logarithm appears in the \order{\alt^2} threshold correction which should be avoided in the EFT approach. Recently, the authors of~\cite{Deppisch:2019iyh} proposed a resummation of terms that are enhanced by powers of the gluino mass as a possibility to alleviate fine-tuning issues in the MSSM and the NMSSM.

In the present paper we demonstrate how state-of-the-art hybrid results that contain a resummation of higher-order logarithmic contributions (the same holds for pure EFT results), can be consistently improved such that large theoretical uncertainties for the case of a heavy gluino are avoided. Our approach is based on the introduction of a suitable renormalisation scheme for the EFT part of the hybrid result, for which the occurrence of power-enhanced corrections from the gluino mass is avoided. We explicitly demonstrate that the terms resummed via the prescription of~\cite{Deppisch:2019iyh} are absorbed by the parameters of the adopted renormalisation scheme. In our numerical analysis we show that reliable theoretical predictions can also be obtained for large hierarchies between the gluino mass and the stop masses. We also discuss the integration of the results into the hybrid framework of the public code \FH~\cite{Heinemeyer:1998yj,Heinemeyer:1998np,Hahn:2009zz, Degrassi:2002fi,Frank:2006yh,Hahn:2013ria,Bahl:2016brp,Bahl:2017aev,Bahl:2018qog}.

The paper is organised as follows. In \cref{sec:02_MGl_resummation}, we discuss how corrections that are enhanced by powers of the gluino mass arise and how they can be absorbed by a suitable choice of the renormalisation scheme. We investigate the numerical implications in \cref{sec:03_results}. Conclusions are given in \cref{sec:04_conclusions}. In the appendix it is shown that our results incorporate the resummation that was proposed in~\cite{Deppisch:2019iyh}.

%%%%%%%%%%%%%%%%%%%%%%%%%%%%%%%%%%%%%%%%%%%%%%%%%%%%%%%%%%%%%%%%%%%%%%%%%%%%%%
%%%%%%%%%%%%%%%%%%%%%%%%%%%%%%%%%%%%%%%%%%%%%%%%%%%%%%%%%%%%%%%%%%%%%%%%%%%%%%

\section{Treatment of contributions enhanced by the gluino mass}
\label{sec:02_MGl_resummation}

In the following we present a systematic approach for the incorporation of terms that can be enhanced by powers of the gluino mass \mgls into the prediction for the mass of the SM-like Higgs boson in the MSSM. We will show that our results automatically incorporate the resummation of large gluino contributions that was recently proposed~\cite{Deppisch:2019iyh}.

In a fixed-order calculation within the OS scheme the leading contributions that are enhanced by powers of the gluino mass cancel out between the unrenormalised diagrams and the counterterms as a consequence of the fact that the OS scheme is a momentum-subtraction scheme. As an example, it is well-known that the unrenormalised self-energies of the scalar top quarks, $\Sigma(p^2)$, receive contributions at the one-loop level that scale proportional to the squared gluino mass in the limit of a heavy gluino. These terms cancel, however, in the renormalised self-energies of the two stop mass eigenstates,
\begin{align}
\label{eq:momsub}
\Sigma^{\rm ren}(p^2) = \Sigma(p^2) - {\rm Re}\left(\Sigma(p^2 = m^2)\right)
+ \ldots ,
\end{align}
where the ellipsis denotes terms involving the field renormalisation constant and $m$ is the mass of the scalar top quark. In the \DR scheme, on the other hand, the mass counterterm does not have a finite part, and a cancellation like in \cref{eq:momsub} does not occur.

In order to treat the case where the gluino is much heavier than the rest of the mass spectrum with EFT methods, the gluino should be integrated out. For this purpose, matching conditions between the full MSSM and the MSSM without gluino have to be calculated. In this matching procedure, all particles except the gluino can be treated as massless. Consequently, it follows purely from dimensional analysis that no terms enhanced by powers of the gluino mass can enter the matching of the Higgs four-point function and also of all other dimensionless Green functions (terms depending logarithmically on the gluino mass are possible).

Contributions that are enhanced by powers of the gluino mass can, however, enter the matching of Green functions with a mass dimension greater than zero. If we perform the matching before electroweak symmetry breaking, these Green functions are all related to soft SUSY-breaking parameters which, apart from the gluino mass parameter $M_3$, can be treated as being zero at the tree-level in the heavy gluino limit. Diagrams involving gluinos generate non-zero contributions at the loop-level which are proportional to powers of the gluino mass. The highest possible power in $M_3$ is given by the mass dimension of the respective parameter.

In the context of the calculation of the lightest SM-like Higgs-boson mass, the soft SUSY-breaking parameters of the scalar top quarks are most relevant.\footnote{The presented argumentation is straightforwardly transferable to other sectors having a smaller numerical impact (e.g.\ the sbottom sector), which are not discussed here. Note that also the Higgs soft-breaking masses receive threshold corrections (see~\cite{Deppisch:2019iyh}). In the present paper, we work in the approximation of setting the electroweak gauge coupling to zero in the non-logarithmic two-loop corrections. Thus, the matching of the Higgs soft-breaking parameters does not enter the calculation of the SM-like Higgs mass.} Their one-loop matching relations (not including terms suppressed by \mgls) read
\begin{align}\label{eq:matching}
& \left(m_{\tilde{t}_{L,R}}^{\MSSM/\tilde g}\right)^2(Q) = \left(m_{\tilde{t}_{L,R}}^\MSSM\right)^2(Q) \left[1 + \frac{\als}{\pi} C_F \frac{\vert M_3 \vert^2}{m_{\tilde{t}_{L,R}}^2}\left(1 + \ln\frac{Q^2}{\vert M_3 \vert^2}\right) \right. - \nonumber \\
& \hspace{3.1cm} - \frac{\als}{4\pi} C_F \left. \left(1 + 2 \ln\frac{Q^2}{\vert M_3 \vert^2} \right)\right], \\
& X_t^{\MSSM/\tilde g}(Q) = X_t^\MSSM(Q) - \frac{\als}{\pi} C_F M_3 \left(1 + \ln\frac{Q^2}{\vert M_3 \vert^2}\right) + \nonumber \\
& \hspace{3.1cm} + \frac{\als}{8\pi} C_F X_t \left(1 - 2 \ln\frac{Q^2}{\vert M_3 \vert^2} \right),
\end{align}
where $\MSSM/\tilde g$ denotes the MSSM without the gluino, $M_3$ is the gluino mass parameter (we consider here the general case where $M_3$ can have complex values), $m_{\tilde{t}_{L,R}}$ are the left and right soft-breaking masses of the stop sector, $X_t$ is the stop mixing parameter, $\als = g_3^2/(4\pi)$ (with $g_3$ being the strong gauge coupling), $C_F = 4/3$ and $Q$ is the scale at which the matching is performed. Higher-order loop corrections to these relations are subleading (i.e., of the form $\vert M_3 \vert^2\als^n$ with $n\ge 2$ in case of the mass parameters and of the form $\vert M_3 \vert \als^n$ in the case of the stop mixing parameter).

After integrating out the gluino at the gluino mass scale, the parameters are evolved down to the stop mass scale, where in the simplest setup all other non-SM particles are integrated out. Since the gluino is not present in the EFT below the gluino mass scale, no terms enhanced by powers of the gluino mass can enter in all the parts of the calculation that are performed at a scale below the gluino mass scale.

\medskip

Deriving all necessary matching conditions for integrating out the gluino as well as the corresponding RGEs is cumbersome.~\footnote{In the recent study~\cite{Kramer:2019fwz} all one-loop matching conditions for operators of dimension four to six were derived for the MSSM without gluino in the gaugeless limit, but in addition also the appropriate two-loop threshold corrections and RGEs would be needed.} Instead, we focus here specifically on terms that are enhanced by powers of the gluino mass. As argued above, these terms arise only in the matching relations of the soft SUSY-breaking masses. Instead of performing the full matching, we can also absorb the terms that are enhanced by powers of the gluino mass into the definition of the parameters. For the case of the mass parameters it is useful to adopt the \MDR scheme employed in \cite{Kant:2010tf} for this purpose,
\begin{align}\label{eq:mst_MDR_def}
\left(m_{\tilde t_{L,R}}^\MDR\right)^2(Q) ={}& \left(m_{\tilde t_{L,R}}^\DR\right)^2(Q) \left[1 + \frac{\als}{\pi} C_F \frac{\vert M_3 \vert^2}{m_{\tilde t_{L,R}}^2}\left(1 + \ln\frac{Q^2}{\vert M_3 \vert^2}\right)\right],
\end{align}
where here $Q$ is the conversion scale at which the \DR parameters are converted to the \MDR ones.

We extend the \MDR scheme by also defining it for the stop mixing parameter,
\begin{align}\label{eq:Xt_MDR_def}
X_t^\MDR(Q) ={}& X_t^\DR(Q) - \frac{\als}{\pi} C_F M_3 \left(1 + \ln\frac{Q^2}{\vert M_3 \vert^2}\right).
\end{align}
Using this scheme, the resummation formulas derived in~\cite{Deppisch:2019iyh} are easily recoverable (see \App{app}). Note also that the \MDR parameters are scale independent at leading order in \mgls.

If the EFT calculation is performed in the \MDR scheme no terms enhanced by powers of the gluino mass appear.\footnote{If three-loop threshold corrections are taken into account, also subleading terms (i.e., terms of two-loop order) in the \MDR definition have to be taken into account~\cite{Martin:2006ub, Kant:2010tf}.} At the same time the occurrence of large logarithms, $\ln\msusy/M_t$, in the threshold corrections, which is not desirable in the EFT approach (as happens for the ``OS-like'' scheme proposed in~\cite{Vega:2015fna}), is avoided in this way. It should be noted that the threshold corrections between the SM and the MSSM still depend logarithmically on the gluino mass. These terms are, however, numerically less problematic.

If \MDR parameters are given as input for the calculation, the application of the \MDR scheme is obviously straightforward. It is, however, desirable to also allow for input parameters renormalised in other schemes. In order to incorporate the EFT calculation using the \MDR scheme into a hybrid result involving on-shell parameters or into a framework using another scheme, for instance \DR parameters, the respective parameters need to be related to the corresponding quantities in the \MDR scheme. We briefly describe in the following how this can be achieved for the case of OS and \DR parameters.

%%%%%%%%%%%%%%%%%%%%%%%%%%%%%%%%%%%%%%%%%%%%%%%%%%%%%%%%%%%%%%%%%%%%%%%%%%%%%%

\subsubsection*{On-shell input parameters}

Often the OS scheme is used for the definition of the stop parameters. It relates the mass parameters directly to physical observables (more precisely, pseudo-observables) and is therefore often used in phenomenological studies.

For the case of OS input parameters the incorporation of the EFT calculation using the \MDR scheme can be carried out along the lines of the procedure that is employed in the hybrid framework of \FH for combining the fixed-order and EFT approaches~\cite{Hahn:2013ria,Bahl:2016brp,Bahl:2017aev,Bahl:2018ykj}. As usual, the fixed-order corrections can be evaluated directly in the OS scheme.

As for the EFT part, there are, in principle, two strategies possible. One can convert the input parameters from the OS to the \MDR scheme at some scale and then use the \MDR quantities obtained in this way in the EFT calculation. However, a logarithm of the SUSY scale over the top mass scale appears in this conversion. As argued in~\cite{Bahl:2016brp,Bahl:2017aev} only this logarithm has to be retained in the formula since it is sufficient to reproduce the logarithms emerging in the fixed-order result. Furthermore, the associated uncertainty from higher-order logarithmic terms is part of the uncertainty estimate presented in~\cite{Bahl:2019hmm}. As an alternative to this method, one could consider using an ``OS-like'' scheme, similar to the one presented in~\cite{Vega:2015fna}, in the EFT calculation. This strategy, however, leads to a large logarithm, $\ln\msusy/M_t$, in the two-loop threshold-correction of the SM Higgs self-coupling. In order to avoid the occurrence of a large logarithm in this threshold correction, we prefer to use the procedure outlined above where such a logarithm appears only in the scheme conversion of the input parameters. To the best of our knowledge it has not yet been shown how logarithms of this kind could be properly resummed. We leave this issue for further study, but as mentioned above include it as part of the estimate of the remaining theoretical uncertainties.

We improve the hybrid result by carrying out the EFT calculation with the stop parameters defined in the \MDR scheme rather than the \DR scheme as it was used up to now (the subtraction terms are adapted accordingly). In order to obtain the input parameters of the EFT calculation in the \MDR scheme we need to convert the input parameters given in the OS scheme to the \MDR scheme. As we already mentioned above, we retain only large one-loop logarithms in this conversion. This implies that the incorporation of the EFT results using the \MDR scheme instead of the \DR scheme does not require changes in the conversion formulas presented in~\cite{Bahl:2016brp}. We perform the conversion at the scale \msusy. The matching scale between the SM and the MSSM can in principle be chosen independently of the conversion scale. We choose to use \msusy. Alternatively, converting at the scale \mgls (and also using \mgls as matching scale) does not lead to large shifts in $M_h$.

%%%%%%%%%%%%%%%%%%%%%%%%%%%%%%%%%%%%%%%%%%%%%%%%%%%%%%%%%%%%%%%%%%%%%%%%%%%%%%

\subsubsection*{\DR input parameters}

For the study of high-scale SUSY breaking models, often the \DR scheme is used as it is appropriate for running down the parameters from the high scale. This is also the scheme that is usually employed in pure EFT calculations. For the case of \DR input parameters the following procedure should be employed. The high-scale \DR parameters are run down to the conversion scale, where they are converted to the \MDR scheme using \cref{eq:mst_MDR_def}. After that the fixed-order as well as the EFT calculation can be carried out in the \MDR scheme.

In principle, the conversion scale can be chosen arbitrarily. The result does not depend on it at the two-loop level. As argued above, the insertion of the \MDR parameters into the one-loop threshold corrections is equivalent to the resummation of $\mgls$-enhanced contributions to all orders. However, this insertion also generates large $\mgls^2$-enhanced logarithmic contributions, see \cref{eq:mst_MDR_def}, unless $Q=\mgls$ is set. Therefore, in our approach, we perform the conversion between the \DR and the \MDR parameters at the gluino mass scale in order to avoid an unstable result. If \DR parameters defined at a scale below \mgls are given as input, an unstable result cannot be avoided. In addition, we also set the scale where the SM is matched to the MSSM, which in principle is independent of the conversion scale, to the scale of the gluino mass.

%%%%%%%%%%%%%%%%%%%%%%%%%%%%%%%%%%%%%%%%%%%%%%%%%%%%%%%%%%%%%%%%%%%%%%%%%%%%%%
%%%%%%%%%%%%%%%%%%%%%%%%%%%%%%%%%%%%%%%%%%%%%%%%%%%%%%%%%%%%%%%%%%%%%%%%%%%%%%

\section{Numerical results}
\label{sec:03_results}

In this Section, we discuss the numerical implications of using the \MDR scheme in the EFT calculation. We focus on a single-scale scenario in which all non-SM mass parameters are chosen to be equal to a common mass scale named \msusy, which is set to 1.5~TeV. As the only exception, we allow the gluino mass parameter $M_3$ to take a different value. We choose all soft SUSY-breaking trilinear couplings to be zero except for the stop trilinear coupling which is fixed by setting the stop mixing parameter $X_t$. The ratio of the Higgs vacuum expectation values, $\tan\beta$, is set to 10.

%%%%%%%%%%%%%%%%%%%%%%%%%%%%%% FIGURE %%%%%%%%%%%%%%%%%%%%%%%%%%%%%%
\begin{figure}
\begin{minipage}{.48\textwidth}\centering
\includegraphics[width=\textwidth]{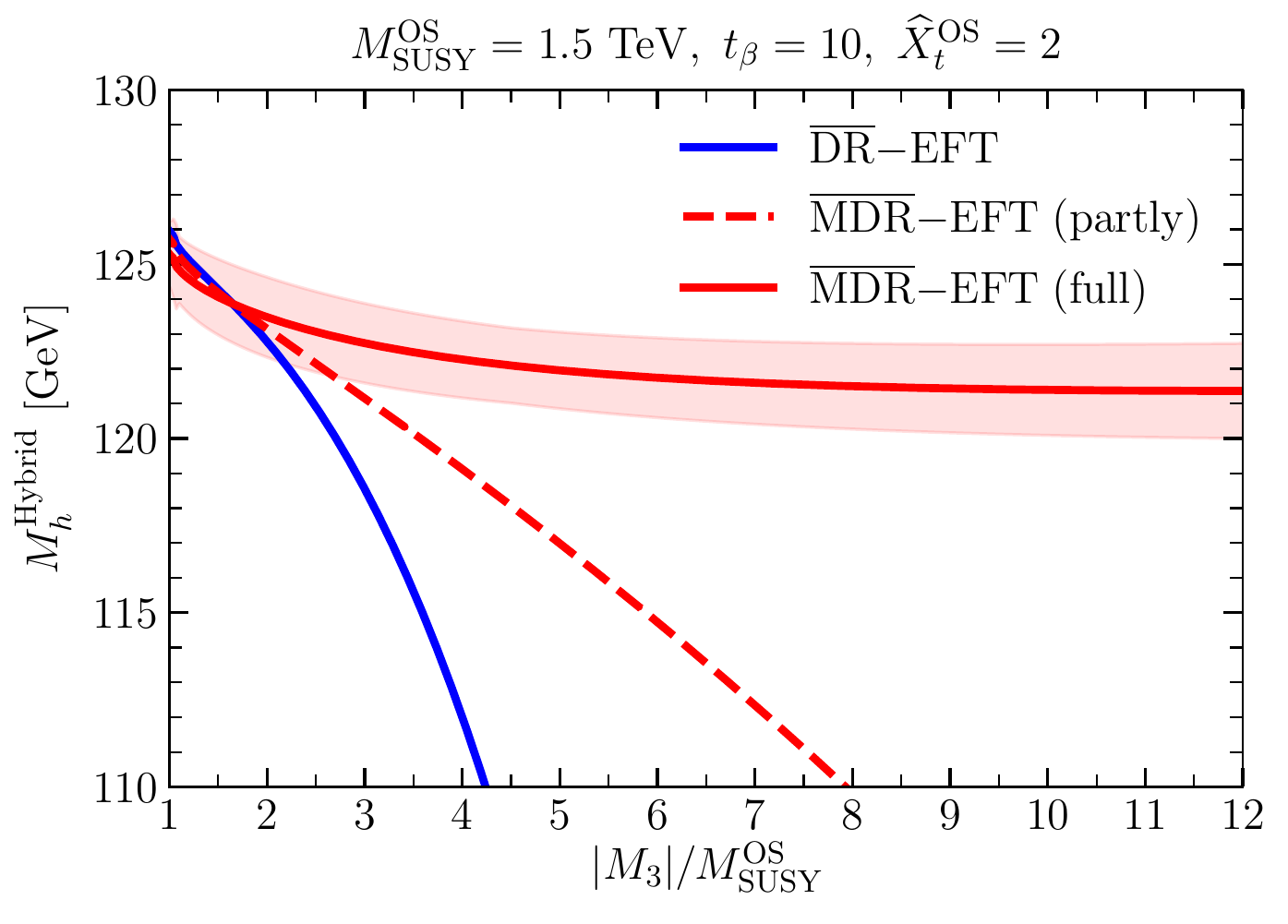}
\end{minipage}
\begin{minipage}{.48\textwidth}\centering
\includegraphics[width=\textwidth]{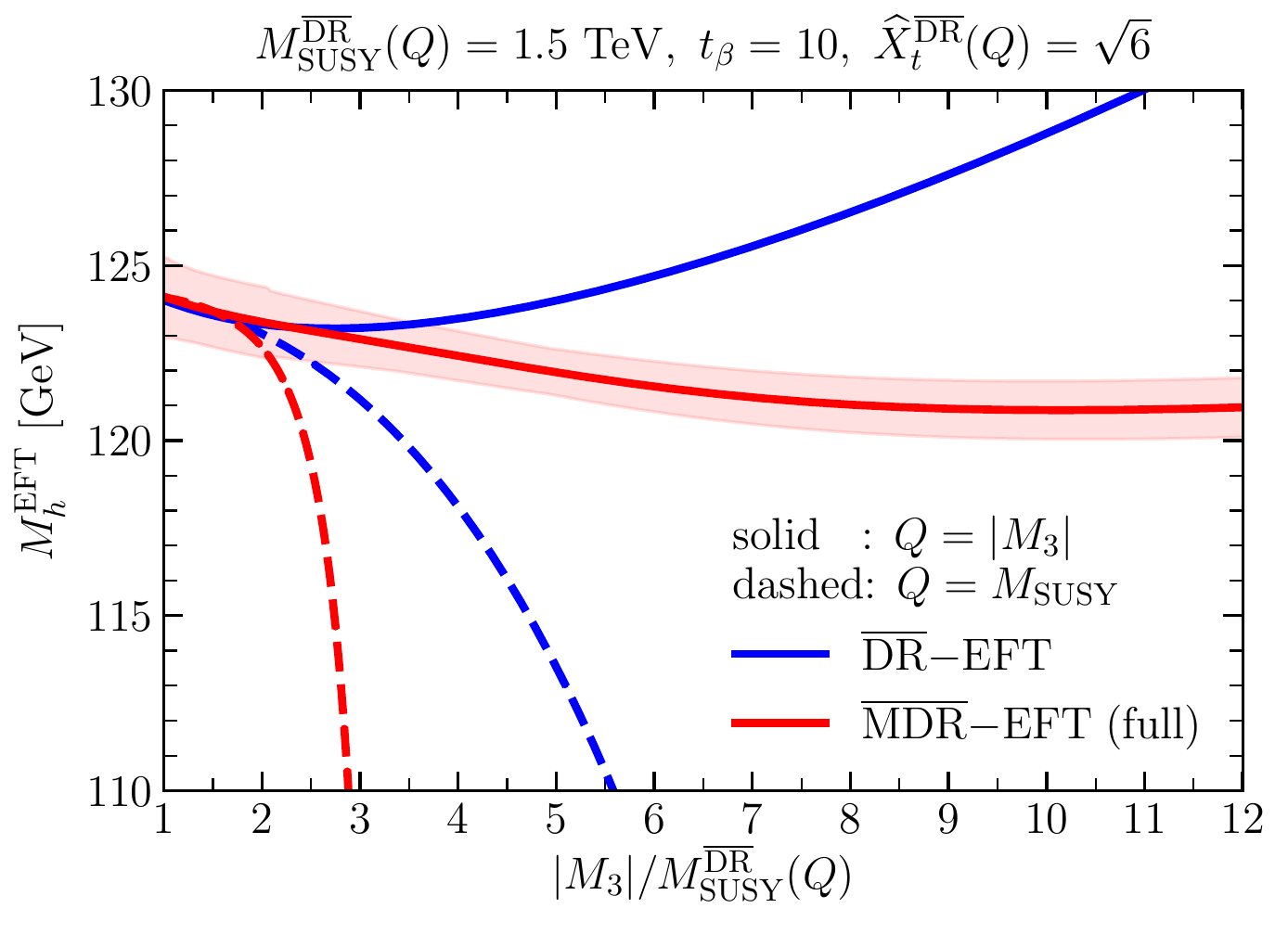}
\end{minipage}
\caption{Prediction for $M_h$ as a function of the ratio of the gluino mass over $\msusy$. For the most accurate result, labelled as \MDR--EFT (full), the estimate of the remaining theoretical uncertainties is shown as coloured band. The result is compared with predictions using different renormalisation schemes in the EFT calculation. \textit{Left}: Prediction of the hybrid calculation using the OS scheme for the definition of the input parameters where for the EFT part the full \MDR scheme, a partial \MDR scheme (see text) and the \DR scheme are used. \textit{Right}: Prediction of the pure EFT calculation using the \DR scheme for the definition of the input parameters. The result for the \MDR scheme with the conversion (and input) scale $Q = |M_3|$ is compared with the one for  $Q = \msusy$ and with the result for the \DR scheme with both scale choices.}
\label{fig:M3scan}
\end{figure}
%%%%%%%%%%%%%%%%%%%%%%%%%%%%%% FIGURE %%%%%%%%%%%%%%%%%%%%%%%%%%%%%%

The left panel of~\cref{fig:M3scan} shows the prediction for the lightest Higgs boson mass in the hybrid approach obtained by \FH (version \texttt{2.15.0}) as a function of the ratio $\mgls/\msusy$. In addition to the \DR scheme, which was used up to now by default in the EFT calculation of \FH for the definition of the stop parameters, we implemented the \MDR scheme as defined in \Sec{sec:02_MGl_resummation}. For this plot the input parameters of the stop sector are assumed to be defined in the OS scheme, setting $\widehat{X}_t^{\rm OS} = 2$, where $\widehat{X}_t = X_t / \msusy$.

The blue solid line shows the result obtained using the \DR scheme in the EFT calculation. This means, in particular, that the stop parameters are \DR parameters defined at the scale \msusy. Since the input parameters are defined in the OS scheme, there is no quadratic and linear dependence on $M_3$ at the two-loop level as the calculation up to this level is based on the fixed-order result in the OS scheme. However, terms that are enhanced by powers of the gluino mass emerge from the two-loop threshold correction to the Higgs quartic coupling of $\mathcal{O}(\alt \als)$. This threshold correction generates three-loop NNLL (next-to-next-to-leading logarithmic) terms, involving \msusy and $M_t$, in the expression for the Higgs-boson mass. For $\widehat{X}_t^{\rm OS} = 2$, these power-enhanced terms in the \DR-EFT result drive the Higgs-mass prediction steeply downwards when $\vert M_3\vert$ increases as shown in the plot. The large numerical impact of the power-enhanced terms leads to a large increase of the theoretical uncertainty of the EFT result in the \DR scheme (and consequently also of the hybrid result into which it is implemented), see the discussion in~\cite{Bahl:2019hmm}. There a drastic increase of the uncertainty was found in the region $\mgls/\msusy \gtrsim 2$ for the same scenario.

The red dashed curve on this plot corresponds to the result of the hybrid calculation in which in the EFT calculation the left and right stop soft-breaking masses are parametrized in the \MDR scheme at the scale \msusy, while the stop mixing parameter $X_t$ is still a \DR parameter defined at the same scale. This corresponds to the version of the \MDR scheme that was previously used in the literature. While for this result the Higgs-mass prediction falls less rapidly with increasing $\vert M_3\vert$ compared to the blue solid curve, the plot shows that there is a remaining approximately linear dependence of the squared Higgs mass on $\vert M_3\vert$ that leads to large theoretical uncertainties also for this result. The reason for the somewhat improved behaviour with respect to the result that is based on the \DR scheme can be traced to the fact that the choice of $m_{\tilde{t}_{L,R}}$ in the \MDR scheme absorbs the quadratic dependence $\sim \vert M_3\vert^2$ in the mentioned three-loop NNLL terms (and also of higher order terms).

The red solid curve in the left plot of~\cref{fig:M3scan} shows the result of the hybrid calculation making use of the extended \MDR scheme as described above. Accordingly, all the stop parameters entering the EFT part of the hybrid result are parametrized in the \MDR scheme at the scale \msusy. In this case, all the terms scaling like powers of the gluino mass that would be induced by the two-loop $\mathcal{O}(\alt \als)$ contribution to the threshold correction are absorbed into the definition of the soft-breaking parameters. We observe only a rather mild logarithmic dependence of the calculated Higgs-boson mass on $\vert M_3\vert/\msusy$. These logarithms could be resummed by performing the complete matching of the full MSSM to the MSSM without a gluino as low-energy theory above the stop mass scale. This, however, is numerically much less relevant and lies beyond the scope of the present paper.

For the full \MDR result, we also show a coloured band indicating the remaining theoretical uncertainty estimated using the procedure developed in~\cite{Bahl:2019hmm}. The comparison with the uncertainty estimate obtained in~\cite{Bahl:2019hmm} for the case where the EFT part of the calculation is based on the \DR scheme shows that the application of the (extended) \MDR scheme to the EFT part of the calculation leads to a drastic reduction of the theoretical uncertainty. The uncertainty that we estimate for the full \MDR--EFT result stays approximately constant ($\sim 1.5\gev$) when \mgls is raised and shows no sharp increase as found in~\cite{Bahl:2019hmm}.

\medskip

The right panel of~\cref{fig:M3scan} shows the lightest Higgs-boson mass calculated in the pure EFT approach as implemented in \FH (version \texttt{2.15.0}) as a function of $\mgls/\msusy$ using the \DR scheme for the definition of the input parameters. We define these parameters at the conversion scale $Q$. For our result based on the (extended) \MDR scheme the parameters have to be defined at the scale $Q =  \mgls$ (red solid line) in order to obtain a numerical stable result (see also \cref{sec:02_MGl_resummation}). For comparison we also show the \MDR--EFT result where the input parameters are defined at \msusy, where also the conversion to the \MDR scheme is carried out (red dashed line). We furthermore display the \DR--EFT result for both input scale choices. It should be noted that for $\mgls \neq \msusy$ the solid lines (input scale \mgls) and the dashed lines (input scale $\msusy$) cannot be directly compared to each other since they represent different physical situations. We fix $\widehat{X}_t^\DR = \sqrt{6}$ for this plot.

Our result based on the (extended) \MDR scheme (red solid line) is parametrized in terms of \MDR quantities at the scale \mgls, which are obtained from a one-loop conversion of the \DR parameters at the same scale using \cref{eq:mst_MDR_def,eq:Xt_MDR_def} and making the according adjustments to the $\mathcal{O}(\alt \als)$ threshold correction (for more details see \Sec{sec:02_MGl_resummation}). We observe only a mild logarithmic dependence of the calculated Higgs-boson mass on $\mgls/\msusy$ for this result.

The coloured band shows the estimated size of unknown higher-order corrections for the \MDR--EFT result with $Q=\mgls$. The theoretical uncertainty of the EFT result is estimated following largely the procedure employed in~\cite{Bahl:2019hmm}. As only differences we take into account the modified scale dependence of the \MDR parameters in comparison to the \DR parameters and an additional uncertainty associated with unknown higher-order corrections to the relations converting the parameters from the \DR to the \MDR scheme (see \cref{eq:mst_MDR_def,eq:Xt_MDR_def}). We estimate this uncertainty by replacing $\als$ by $\als\left[1 \pm \als/(4\pi)\left(1 + \ln Q^2 / \mgls^2\right)\right]$ in \cref{eq:mst_MDR_def,eq:Xt_MDR_def} (see also the discussion in~\cite{Bahl:2019hmm}). As for the case of OS input parameters for the hybrid result, the total uncertainty stays approximately constant ($\sim 1\gev$) when \mgls is raised.

For comparison, the red dashed line shows the EFT result based on the (extended) \MDR scheme where the scale of the \DR input parameters is chosen as $Q=\msusy$ instead of $Q =  \mgls$. It is clearly visible that using \DR parameters defined at the scale \msusy as input spoils the stability of the \MDR--EFT result. Numerically, the sharp decrease of $M_h$ is largely driven by the behaviour of the \MDR parameters. As one can see from \cref{eq:mst_MDR_def}, in this scenario, the \MDR stop soft-breaking masses decrease with increasing \mgls while the stop mixing parameter $X_t^\MDR(Q)$ increases. This results in the suppression of $M_h$ with increasing \mgls visible for the dashed red line. In contrast, the \MDR stop soft-breaking masses increase and $X_t^\MDR(Q)$ decreases for $Q =  \mgls$ (red solid line) resulting in the observed stability for rising \mgls.\footnote{This is true for the given choice of $M_3$ (i.e., $M_3>0$). For $M_3<0$ (or a complex-valued $M_3$), $|X_t^\MDR(\mgls)|$ can increase but the ratio $\vert \xt \vert  = \vert X_t/\msusy \vert$ will still decrease.} As argued in \cref{sec:02_MGl_resummation}, the instability for $Q = \msusy$ is a consequence of the input parameters being defined in the MSSM and not in a valid EFT in which the gluino is integrated out at the scale \mgls.

We now turn to the discussion of the \DR--EFT result (blue solid and dashed lines). It is obvious that neither input scale choice yields a reliable theoretical prediction of the \DR--EFT calculation. As explained above, this is caused by the power-enhanced gluino contributions that are present in this result. The blue dashed line shows the result where the input scale is chosen as \msusy. In this case, the two-loop threshold correction of $\mathcal{O}(\alt \als)$ depends quadratically on \mgls and so does the squared Higgs mass which contains terms proportional to $\sim \mgls^2 \left(1 + \log \msusy^2/\mgls^2\right)$.

The blue solid curve in the right plot of~\cref{fig:M3scan} corresponds to the \DR--EFT result where the \DR parameters are defined at the scale \mgls. As explained above, there is a quadratic dependence on the gluino mass in the two-loop threshold correction. However, since for the choice $Q = \mgls$ there is no additional negative logarithmic contribution to the threshold correction, in this case the Higgs mass grows with increasing \mgls.

%%%%%%%%%%%%%%%%%%%%%%%%%%%%%%%%%%%%%%%%%%%%%%%%%%%%%%%%%%%%%%%%%%%%%%%%%%%%%%
%%%%%%%%%%%%%%%%%%%%%%%%%%%%%%%%%%%%%%%%%%%%%%%%%%%%%%%%%%%%%%%%%%%%%%%%%%%%%%

\section{Conclusions}
\label{sec:04_conclusions}

In this paper, we have shown how an appropriate choice of the renormalisation prescription for the stop sector of the MSSM leads to a very significant improvement of the theoretical prediction for the mass of the SM-like Higgs boson in the region where the gluino is heavier than the scalar top quarks. This region is phenomenologically important in particular in view of the strengthened limits from experimental searches for the gluino.

In pure EFT calculations and in the EFT part of hybrid results making use of the \DR scheme for the renormalisation of the stop sector leads to the appearance of terms enhanced by powers of the gluino mass. These large non-decoupling effects of the gluino, which are formally of three-loop order for a hybrid result where the fixed-order part is evaluated in the OS scheme up to the two-loop order, lead to unreliable predictions and correspondingly large theoretical uncertainties in the heavy-gluino region. We have shown how the occurrence of power-enhanced corrections from the gluino mass in the EFT part of the calculations can be avoided without spoiling the underlying assumptions of the EFT. In fact, we have demonstrated that the leading contributions from integrating out the gluino can be taken into account by absorbing them into the renormalisation of the stop parameters. This scheme, called \MDR, has already been used before in the literature. We have extended it to include also the stop mixing parameter. We have furthermore shown that the recently proposed resummation of large gluino contributions~\cite{Deppisch:2019iyh} is taken into account in the (extended) \MDR scheme via the absorption of the contributions into the parameters of the model. We also discussed the implementation of the (extended) \MDR scheme into the public code \FH and its impact on the estimate of the remaining theoretical uncertainties from unknown higher-order corrections. The implementation will be publicly released in an upcoming version.

In our numerical analysis we have demonstrated that using the (extended) \MDR scheme for the EFT part of the hybrid result leads to a prediction for $M_h$ that shows only a mild dependence on the gluino mass even for large hierarchies between the gluino mass and the stop masses. The theoretical uncertainties in the heavy-gluino region are vastly improved compared to the case where the EFT result is based on the \DR scheme. We have demonstrated that these features only hold for the extended \MDR scheme, while restricting the scheme to the masses --- as previously used in the literature --- would not be sufficient for this purpose. We have furthermore stressed that in the case of \DR input parameters the scale of the input parameters has to be $|M_3|$ (or larger) in order to allow for a stable $M_h$ prediction for the case where $|M_3| > \msusy$. We have also pointed out that for the EFT approach using the \DR scheme neither the input scale $|M_3|$ nor $\msusy$ leads to a reliable prediction in the heavy-gluino region.

The presented renormalisation prescription is straightforwardly applicable to the calculation of other observables.

%%%%%%%%%%%%%%%%%%%%%%%%%%%%%%%%%%%%%%%%%%%%%%%%%%%%%%%%%%%%%%%%%%%%%%%%%%%%%%
%%%%%%%%%%%%%%%%%%%%%%%%%%%%%%%%%%%%%%%%%%%%%%%%%%%%%%%%%%%%%%%%%%%%%%%%%%%%%%

\section*{Acknowledgements}
\sloppy{
We thank S.~Heinemeyer, W.~Hollik and P.~Slavich for interesting discussions. We acknowledge support by the Deutsche Forschungsgemeinschaft (DFG, German Research Foundation) under Germany's Excellence Strategy -- EXC 2121 ``Quantum Universe'' – 390833306. H.B.\ and G.W.\ thank the Max-Planck-Institut f\"ur Physik, Munich, for hospitality during the final stages of this work.
}

%%%%%%%%%%%%%%%%%%%%%%%%%%%%%%%%%%%%%%%%%%%%%%%%%%%%%%%%%%%%%%%%%%%%%%%%%%%%%%
%%%%%%%%%%%%%%%%%%%%%%%%%%%%%%%%%%%%%%%%%%%%%%%%%%%%%%%%%%%%%%%%%%%%%%%%%%%%%%
%%%%%%%%%%%%%%%%%%%%%%%%%%%%%%%%%%%%%%%%%%%%%%%%%%%%%%%%%%%%%%%%%%%%%%%%%%%%%%

\appendix

%%%%%%%%%%%%%%%%%%%%%%%%%%%%%%%%%%%%%%%%%%%%%%%%%%%%%%%%%%%%%%%%%%%%%%%%%%%%%%
%%%%%%%%%%%%%%%%%%%%%%%%%%%%%%%%%%%%%%%%%%%%%%%%%%%%%%%%%%%%%%%%%%%%%%%%%%%%%%

\section{From the \MDR scheme to the resummation of gluino contributions}
\label{app}

Here, we discuss how the resummation formulas given in \cite{Deppisch:2019iyh} can be recovered from the expressions in the \MDR scheme. The authors of~\cite{Deppisch:2019iyh} considered two sets of diagrams contributing to the matching of the soft SUSY-breaking Higgs mass parameter $m_{22}$.

The left diagram of Fig.~1 in~\cite{Deppisch:2019iyh} corresponds to an $A_0$ Passarino--Veltman loop function. In the \MDR scheme, we do not have to consider any stop self-energy insertions beyond the one-loop diagram. Following~\cite{Deppisch:2019iyh}, we define $\xi_{L,R}$ via
\begin{align}
(m_{\tilde{t}_{L,R}}^\MDR)^2 = (m_{\tilde{t}_{L,R}}^\DR)^2 \left(1 - \xi_{L,R}\right)
\end{align}
we obtain for the finite part of the diagram (with $\mu_R$ being a generic renormalization scale)
\begin{align}
& A_0^\text{fin}\left((m_{\tilde{t}_{L,R}}^\MDR)^2\right) ={}\nonumber\\
{} &= A_0^\text{fin}\left((m_{\tilde{t}_{L,R}}^\DR)^2\right) - (m_{\tilde{t}_{L,R}}^\DR)^2\left(-\xi_{L,R} \ln\frac{(m_{\tilde{t}_{L,R}}^\DR)^2}{\mu_R^2} + \frac{1}{2}\xi_{L,R}^2 + \frac{1}{6}\xi_{L,R}^3 + \dots\right) \nonumber\\
{} &= A_0^\text{fin}\left((m_{\tilde{t}_{L,R}}^\DR)^2\right) - (m_{\tilde{t}_{L,R}}^\DR)^2\left(-\xi_{L,R} \ln\frac{(m_{\tilde{t}_{L,R}}^\DR)^2}{\mu_R^2} + \sum_{k=2}^\infty \frac{\xi_{L,R}^k}{k(k-1)}\right) \nonumber\\
{} &= A_0^\text{fin}\left((m_{\tilde{t}_{L,R}}^\DR)^2\right) - (m_{\tilde{t}_{L,R}}^\DR)^2\left(-\xi_{L,R} \ln\frac{(m_{\tilde{t}_{L,R}}^\DR)^2}{\mu_R^2} + \xi_{L,R} +  (1 - \xi_{L,R})\ln(1 - \xi_{L,R}) \right),
\end{align}
recovering the resummation in Eqs.~(9) and (10) of \cite{Deppisch:2019iyh}.\footnote{The additional two-loop terms found in Eq.~(9) of~\cite{Deppisch:2019iyh} originate from the matching of the soft SUSY-breaking Higgs mass parameter $m_{22}$ and are not related to the matching of the soft SUSY-breaking stop masses. This additional contribution, however, does not affect the calculation of the SM-like Higgs mass at the considered order.}

The right diagram of Fig.~1 in~\cite{Deppisch:2019iyh} corresponds to a $B_0$
Passarino--Veltman loop function. Analogously to the $A_0$ loop function
above, we obtain for the special case where
$m_{\tilde{t}_{L}}^\MDR = m_{\tilde{t}_{R}}^\MDR$.
\begin{align}
B_0\left(0,(m_{\tilde{t}_{L}}^\MDR)^2,(m_{\tilde{t}_{R}}^\MDR)^2\right) &={}B_0\left(0,(m_{\tilde{t}_{L}}^\DR)^2,(m_{\tilde{t}_{R}}^\DR)^2\right) + \xi_{L,R}  + \frac{1}{2}\xi_{L,R}^2 + \frac{1}{3}\xi_{L,R}^3 + \dots \nonumber\\
{} &= B_0\left(0,(m_{\tilde{t}_{L}}^\DR)^2,(m_{\tilde{t}_{R}}^\DR)^2\right) + \sum_{k=1}^\infty \frac{\xi_{L,R}^k}{k} \nonumber\\
{} &= B_0\left(0,(m_{\tilde{t}_{L}}^\DR)^2,(m_{\tilde{t}_{R}}^\DR)^2\right) - \ln(1 - \xi_{L,R}) ,
\end{align}
recovering the expression given in Eq.~(11) of~\cite{Deppisch:2019iyh}.

%%%%%%%%%%%%%%%%%%%%%%%%%%%%%%%%%%%%%%%%%%%%%%%%%%%%%%%%%%%%%%%%%%%%%%%%%%%%%%
%%%%%%%%%%%%%%%%%%%%%%%%%%%%%%%%%%%%%%%%%%%%%%%%%%%%%%%%%%%%%%%%%%%%%%%%%%%%%%

%%%%%%%%%%%%%%%%%%%%%%%%%%%%%%%%%%%%%%%%%%%%%%%%%%%%%%%%%%%%%%%%%%%%%%%%%%%%%%
%%%%%%%%%%%%%%%%%%%%%%%%%%%%%%%%%%%%%%%%%%%%%%%%%%%%%%%%%%%%%%%%%%%%%%%%%%%%%%
%%%%%%%%%%%%%%%%%%%%%%%%%%%%%%%%%%%%%%%%%%%%%%%%%%%%%%%%%%%%%%%%%%%%%%%%%%%%%%

\newpage

\bibliographystyle{JHEP.bst}
\bibliography{bibliography}{}

\end{document}